\documentclass[12pt,draftcls,onecolumn,journal]{IEEEtran}
\usepackage{amsfonts}
\usepackage{amsfonts}
\usepackage{setspace}
\usepackage{clrscode}

\usepackage{amsmath}
\usepackage{amssymb}
\usepackage[dvips]{graphicx}
\usepackage{epsfig}
\usepackage{hyperref}

\usepackage{amsmath}
\usepackage{amssymb}
\usepackage[dvips]{graphicx}
\usepackage{epsfig}

\newcommand{\tsnr}{{\text{\footnotesize{SNR}}}}

\newcommand{\E}{\mathbb{E}}

\newcommand{\C}{{\sf{C}}}

\newcommand{\Pb}{\bar{P}}

\newcommand{\figsize}{0.65}

\newtheorem{Lem1}{Proposition}
\newtheorem{Lem}{Theorem}

\newtheorem{Def}{Definition}

\begin{document}

%
\title{Secure Wireless Communication and Optimal Power Control under Statistical Queueing Constraints}



%
\author{
\vspace{1cm}
\authorblockN{Deli Qiao, Mustafa Cenk Gursoy, and Senem
Velipasalar}
\thanks{The authors are with the Department of Electrical
Engineering, University of Nebraska-Lincoln, Lincoln, NE, 68588
(e-mails: dqiao726@huskers.unl.edu, gursoy@engr.unl.edu,
velipasa@engr.unl.edu).}
\thanks{This work was supported by the National Science Foundation under Grants CNS -- 0834753, and CCF -- 0917265. The material in
this paper was presented in part at the the IEEE International Symposium on Information Theory (ISIT), Austin, TX, in June 2010.}}


\maketitle

\begin{abstract}
In this paper, secure transmission of information over
fading broadcast channels is studied in the presence of statistical
queueing constraints. 
Effective capacity is employed as a performance metric 
 to identify the secure
throughput of the system, i.e., \emph{effective secure throughput}.
It is assumed that perfect channel side information (CSI) is
available at both the transmitter and the receivers. Initially, the scenario in which the transmitter sends common messages to two receivers and confidential messages to one receiver is considered. For this case, \emph{effective secure throughput region}, which is the region of constant arrival rates of common and confidential messages that can be supported by the buffer-constrained transmitter and fading broadcast channel, is defined. It is proven that this
effective throughput region is convex. Then, the optimal power control
policies that achieve the boundary points of the effective
secure throughput region are investigated and an algorithm for the numerical computation of the optimal power adaptation schemes is provided. Subsequently, the special case in which
the transmitter sends only confidential messages to one receiver, is addressed in more detail. For this case,  effective secure throughput is formulated and two different power adaptation policies are studied. In particular, it is noted that 
opportunistic transmission is no longer optimal under buffer constraints and the transmitter
should not wait to send the data at a high rate until the main
channel is much better than the eavesdropper channel. 
\end{abstract}

\begin{spacing}{1.8}
\section{Introduction}
Security is an important consideration in wireless systems due to the
broadcast nature of wireless transmissions. In a pioneering work,
Wyner in \cite{wyner} addressed the security problem from an
information-theoretic point of view and considered a wiretap
channel model. He proved that secure transmission of confidential
messages to a destination in the presence of a degraded wire-tapper
can be achieved, and he established the secrecy capacity which is
defined as the highest rate of reliable communication from the
transmitter to the legitimate receiver while keeping the wire-tapper
completely ignorant of the transmitted messages. Recently, there has
been numerous studies addressing information theoretic security
\cite{lai}-\cite{ruoheng}. For instance, the impact of fading has
been investigated in \cite{lai}, where it has been shown that a
non-zero secrecy capacity can be achieved even when the eavesdropper
channel is better than the main channel on average. The secrecy
capacity region of the fading broadcast channel with confidential
messages and associated optimal power control policies have been
identified in \cite{liangsecure}, where it is shown that the
transmitter allocates more power as the strength of the main channel
increases with respect to that of the eavesdropper channel.

In addition to security issues, providing acceptable performance and
quality is vital to many applications. For instance, voice over IP
(VoIP), interactive-video (e.g,. videoconferencing), and streaming-video systems are
required to satisfy certain buffer or delay constraints, and the recent proliferation and expected widespread use of multimedia applications in next generation wireless systems call for a rigorous performance analysis under such quality of service (QoS) considerations. A performance measure for these systems is the effective capacity \cite{dapeng}, which can be seen as the maximum constant arrival
rate that a given time-varying service process can support while
satisfying statistical QoS constraints imposed in the form of limitations on the buffer length. Effective capacity is recently studied in various wireless scenarios (see e.g., \cite{jia}--\cite{fixed} and
references therein). For instance,  Tang and Zhang in \cite{jia}
considered the effective capacity when both the receiver and
transmitter know the instantaneous channel gains, and derived the
optimal power and rate adaptation policies that maximize the
system throughput under QoS constraints. Liu \emph{et al.} in
\cite{finite} considered fixed-rate transmission schemes and
analyzed the effective capacity and related resource requirements
for Markov wireless channel models. In \cite{deli} and \cite{fixed}, energy efficiency is addressed when the wireless systems operate under buffer constraints and employ either adaptive or fixed transmission schemes.

The above-mentioned studies addressed the physical-layer security and QoS limitations separately. However, the joint treatment of these considerations is of much interest from both practical and theoretical points of view. The practical relevance is through, for instance, the wide range of military and commercial applications and scenarios in which sensitive multimedia information needs to be transmitted in a wireless and secure fashion. The theoretical interest is due to the certain tension that arises when both secrecy and buffer limitations are present. For instance, physical layer security leads to lower transmission rates. Moreover, the optimal performance in wireless scenarios requires opportunistic transmissions in which one has to wait for high-rate transmission until the main channel between the transmitter and the legitimate receiver is much stronger than the eavesdropper's channel. Note that both end-results may cause buffer overflows and packet losses and may be detrimental in buffer/delay constrained systems. Despite these motivating facts,
the combination of security and delay/buffer considerations has received only little attention so far.  In
\cite{liangqueue}, Liang \emph{et al.}
analyzed the arrival
rates supported by a fading wire-tap channel and identified the power allocation policies that take into
account the queue lengths. In \cite{delaysecrecy}, Youssef \emph{et al.} studied the delay limited secrecy capacity of fading channels.

In this paper, we address both physical-layer security issues and buffer limitations in order to identify the key tradeoffs and optimal transmission strategies. We
assume that perfect channel side information (CSI) is available at both the
transmitter and receivers. We first consider a secure broadcasting scenario in which the transmitter sends common messages to two receivers and confidential messages to one receiver. For this case, we define the \emph{effective
secrecy throughput region} as the region of common and confidential message arrival rates
that can be supported when the transmitter operates under constraints on buffer violation probabilities. Then, we investigate the optimal power
allocation policies that achieve points on the boundary of the
effective secrecy throughput region. We provide an algorithm to determine
the power allocation as a function of  the channel states. Subsequently, we provide a more detailed analysis of the special case in which the transmitter sends no common messages. In this case, the broadcast channel with confidential messages is reduced to
a wiretap channel. In this
scenario, we provide the expression for the effective secure throughput  and analyze two types of control policies by adapting the
power with respect to both the main and eavesdropper channel
conditions and also with respect to only the main channel
conditions. Through this analysis, we find that, due to the
introduction of the buffer constraints, the transmitter cannot reserve
its power for times at which the main channel is much stronger than
the eavesdropper channel. Also, we find that adapting the power
allocation strategy with respect to both the main and eavesdropper
channel CSI rather than only the main channel CSI provides little
improvement when QoS constraints become more stringent.

The rest of the paper is organized as follows. Section II briefly
describes the system model and the necessary preliminaries on
statistical QoS constraints and effective capacity. In Section III,
the effective secrecy throughput region and the corresponding
optimal power allocation policies are presented for fading broadcast channels with confidential messages. In Section IV, the
special case in which the common message rate is zero is studied in more detail. Finally,
Section V concludes the paper.

\section{System Model and Preliminaries}

\begin{figure}
\begin{center}
\includegraphics[width=\figsize\textwidth]{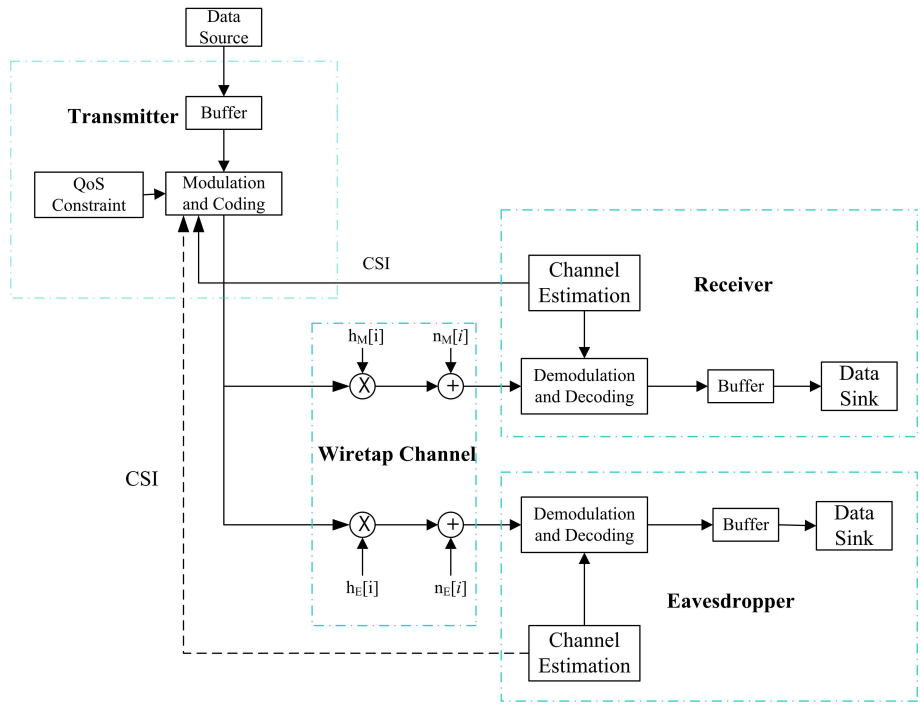}
\caption{The general system model.}\label{fig:systemmodel}
\end{center}
\end{figure}
\subsection{System and Channel Models}

As depicted in Figure \ref{fig:systemmodel}, we consider a system with one transmitter and two receivers. We assume that the transmitter sends confidential messages to receiver 1. From this perspective, receiver 2 can be regarded as an eavesdropper. However, receiver 2 is not necessarily a malicious eavesdropper as we also consider a broadcast scenario in which transmitter sends common messages to both receivers.

In the model, data sequences generated by the source
are divided into frames of duration $T$. These data frames are
initially stored in the buffer before they are transmitted over the
wireless channel. The channel input-output relationships are given by
\begin{align}
Y_1[i]=h_1[i]X[i]+W_1[i] \intertext{and} Y_2[i]=h_2[i]X[i]+W_2[i]
\end{align}
where $i$ is the symbol index, $X[i]$ is the channel input in the
$i$th symbol duration, and $Y_1[i]$ and $Y_2[i]$ represent the
channel outputs at receivers 1 and 2, respectively. We assume
that $\{h_j[i],\, j=1,2\}$'s are jointly stationary and ergodic
discrete-time processes, and we denote the magnitude-square of the
fading coefficients by $z_j[i]=|h_j[i]|^2$. Considering the
receiver 1 as the main user, to which we send both the common and
confidential messages, and regarding receiver 2 as the eavesdropper for the
confidential messages,  we replace $z_1$ with $z_M$ and $z_2$ with
$z_E$ to increase the clarity in the subsequent formulations.  The
channel input is subject to an average energy constraint
$\E\{|X[i]|^2\} \le \Pb/B$ where $B$ is the bandwidth available
in the system and hence $\Pb$ is average power constraint (under the assumption that the symbol rate is $B$ complex symbols per second). Above, $W_j[i]$ is a zero-mean, circularly
symmetric, complex Gaussian random variable with variance
$\E\{|W_j[i]|^2\} = N_j$. The additive Gaussian noise samples
$\{W_j[i]\}$ are assumed to form an independent and identically
distributed (i.i.d.) sequence.

We denote the average transmitted signal to noise ratio
with respect to receiver 1 as $\tsnr=\frac{\Pb}{N_1 B}$. Also, we denote the instantaneous transmit power in the $i$th frame as $P[i]$. Now, the
instantaneous transmitted SNR level for receiver 1 becomes
$\mu^1[i]=\frac{P[i]}{N_1 B}$. Then, the average power constraint at
the transmitter is equivalent to the average SNR constraint
$\E\{\mu^1[i]\}\le \tsnr$ for receiver 1 \cite{book}. If we denote
the ratio between the noise powers of the two channels as
$\gamma=\frac{N_1}{N_2}$, the instantaneous transmitted SNR level
for receiver 2 becomes $\mu^2[i]=\gamma\mu^1[i]$.

\subsection{Secrecy Capacity Region with Common Messages} \label{subsec:capacityregion}

We consider a block-fading channel in which the fading coefficients stay constant for the block duration of $T$ seconds and change independently across the blocks. We assume that both the transmitter and receivers have perfect channel side information (CSI). Equipped with the channel knowledge, the transmitter employs power control. We denote
the power allocation policies for the common and confidential messages by $\mu=(\mu_0(\mathbf{z}),\mu_1(\mathbf{z}))$, respectively, where $\mathbf{z}=(z_M,z_E)$ is the vector composed of the
channel states of receivers 1 and 2. Note that the power control policies are defined as instantaneous power levels normalized by the noise power $N_1$ at receiver 1, i.e., $\mu_0(\mathbf{z}) = \frac{P_0[i]}{N_1 B}$ and $\mu_1(\mathbf{z}) = \frac{P_1[i]}{N_1 B}$ where $P_0[i]$ and $P_1[i]$ are the instantaneous powers of the common and confidential messages as functions of the fading states $\mathbf{z}$. The region of fading states in which confidential messages are transmitted is $\mathcal{Z}=\left\{\mathbf{z} \ge 0: \, z_M>\gamma
z_E\right\}$ while the complement of this region in the first quadrant
is  $\mathcal{Z}^c = \left\{\mathbf{z} \ge 0: \, z_M\le\gamma z_E \right\}$. In $\mathcal{Z}^c$, eavesdropper's channel is stronger and instantaneous secrecy capacity is zero. Hence, when $\mathbf{z}\in\mathcal{Z}^c$, confidential messages are not transmitted and
$\mu_1(\mathbf{z})=0$. Following the above definitions, we finally
define $\mathcal{U}$ as set of the power allocation policies that
satisfy the average  $\tsnr$ constraint, i.e.,
\begin{align}\label{eq:avgpower}
\mathcal{U}=\left\{\mu:\, \E_{\mathbf{z}\in\mathcal{Z}
}\{\mu_0(\mathbf{z})+\mu_1(\mathbf{z})\}+\E_{\mathbf{z}\in\mathcal{Z}^c}\{\mu_0(\mathbf{z})\}\le\tsnr = \frac{\Pb}{N_1 B}\right\}.
\end{align}

With the above power control policies, the maximum instantaneous
common message rate in each block with power control policy $\mu$ is given by \cite[Section V]{liangsecure}
\begin{align}\label{eq:commonrate}
R_0=\left\{\begin{array}{ll}\log_2(1+\frac{\gamma\mu_0(\mathbf{z})z_E}{1+\gamma
\mu_1(\mathbf{z})z_E}),&\mathbf{z}\in
\mathcal{Z}\\
\log_2(1+\mu_0(\mathbf{z})z_M),& \mathbf{z}\in\mathcal{Z}^c
\end{array}\right.
\end{align}
under the assumption that channel coding is performed in each block of duration $T$ seconds, and the block length, which is $TB$ symbols, is large enough so that the probability of error is negligible and hence communication at these rates is reliable. Under similar assumptions, the maximum instantaneous confidential message rate is given by
\begin{align}\label{eq:inst-secrecyrate}
R_1=\left\{
\begin{array}{ll}
\log_2\left(1+\mu_1(\mathbf{z})z_M\right)-\log_2\left(1+\gamma\mu_1(\mathbf{z})z_E\right) &\mathbf{z}\in
\mathcal{Z}\\
0,& \mathbf{z}\in\mathcal{Z}^c
\end{array}\right.
\end{align}
Then, the ergodic secrecy capacity region for the fading broadcast channel with common and confidential messages is\footnote{Note that if coding over all channel channel states is allowed, a larger capacity region can be achieved (see e.g., \cite[Section IV]{liangsecure}).}
\begin{align}
\mathcal{C}_s=\bigcup_{\mu\in\mathcal{U}}\left\{
\begin{array}{l}
(R_{0,avg},R_{1,avg}):\\
R_{0,avg}\le\E_{\mathbf{z}\in\mathcal{Z}}\left\{\log_2(1+\frac{\gamma\mu_0(\mathbf{z})z_E}{1+\gamma
\mu_1(\mathbf{z})z_E})\right\}
+\E_{\mathbf{z}\in\mathcal{Z}_c}\left\{\log_2(1+\mu_0(\mathbf{z})z_M)\right\}\\
R_{1,avg}\le\E_{\mathbf{z}\in\mathcal{Z}}\left\{\log_2\left(1+\mu_1(\mathbf{z})z_M\right)-\log_2\left(1+\gamma\mu_1(\mathbf{z})z_E\right)\right\}
\end{array}
\right\}.
\end{align}
The following result shows the convexity of the above capacity region.
\begin{Lem1} \label{prop:convexity-ergodic}
The ergodic secrecy capacity region $\mathcal{C}_s$ is convex.
\end{Lem1}
\emph{Proof: }Let $\mathbf{R} = (R_{0,avg}, R_{1,avg}) \in\mathcal{C}_s  \text{ and } \mathbf{R}' = (R'_{0,avg}, R'_{1,avg}) \in\mathcal{C}_s$ be
two rate pairs achieved by power control policies
$\mu=(\mu_0(\mathbf{z}),\mu_1(\mathbf{z})) \in\mathcal{U}$ and
$\mu'=(\mu'_0(\mathbf{z}),\mu'_1(\mathbf{z})) \in\mathcal{U}$, respectively. Now, assume that a
time-sharing strategy is employed, and power control policy $\mu$ is used $\alpha \in (0,1)$ fraction of the time and $\mu'$ is employed in the remaining $1-\alpha$ fraction of the time. The new power control policy can be expressed as
\begin{gather}
\mu^*=\left\{\begin{array}{ll} \mu, &\text{$\alpha$ fraction of the time}
\\
\mu',& \text{$1-\alpha$ fraction of the time}
\end{array}\right..
\end{gather}
Since $\mu \in \mathcal{U}$ and $\mu' \in\mathcal{U}$, we can easily see that
$\E\{\mu^*\}=\alpha\E\{\mu\}+(1-\alpha)\E\{\mu'\}\le\tsnr$, and hence $\mu^* \in\mathcal{U}$ as well. Moreover, this time-sharing strategy achieves
$\alpha \mathbf{R}+(1-\alpha)\mathbf{R}'$ with the power control policy $\mu^*$. Therefore, we conclude that $\mathbf{R}+(1-\alpha)\mathbf{R}' \in \mathcal{C}_s$, showing the convexity of $\mathcal{C}_s$. \hfill$\square$

\subsection{Statistical QoS Constraints and Effective Secure Throughput}
In \cite{dapeng}, effective capacity is defined as the
maximum constant arrival rate\footnote{For time-varying arrival
rates, effective capacity specifies the effective bandwidth of the
arrival process that can be supported by the channel.} that a given
service process can support in order to guarantee a statistical QoS
requirement specified by the QoS exponent $\theta$. If we define $Q$
as the stationary queue length, then $\theta$ is the decay rate of
the tail distribution of the queue length $Q$:
\begin{equation}
\lim_{q \to \infty} \frac{\log P(Q \ge q)}{q} = -\theta.
\end{equation}
Therefore, for large $q_{\max}$, we have the following approximation
for the buffer violation probability: $P(Q \ge q_{\max}) \approx
e^{-\theta q_{\max}}$. Hence, while larger $\theta$ corresponds to
more strict QoS constraints, smaller $\theta$ implies looser QoS
guarantees. Similarly, if $D$ denotes the steady-state delay
experienced in the buffer, then $P(D \ge d_{\max}) \approx
e^{-\theta \delta d_{\max}}$ for large $d_{\max}$, where $\delta$ is
determined by the arrival and service processes
\cite{tangzhangcross2}.

The effective capacity is given by
\begin{align}\label{eq:effectivedefi}
C(\theta)=-\lim_{t\rightarrow\infty}\frac{1}{\theta
t}\log_e{\mathbb{E}\{e^{-\theta S[t]}\}} \,\quad \text{bits/s},
\end{align}
where the expectation is with respect to $S[t] =
\sum_{i=1}^{t}s[i]$, which is the time-accumulated service process.
$\{s[i], i=1,2,\ldots\}$ denote the discrete-time stationary and
ergodic stochastic service process. We define the effective capacity
obtained when the service rate is confined by the secrecy capacity
region as the \emph{effective secure throghput}.

Under the block fading assumption, the service rate in the $i$th block is $s[i]=TB R$ bits per $T$ seconds, where
$R$ is the instantaneous service rate for either common or confidential
messages, and $B$ is the bandwidth. Note that $s[i]$ varies independently from one block to another due to the block fading assumption. Then,
(\ref{eq:effectivedefi}) can be written as
\begin{align}
C(\theta)&=
-\frac{1}{\theta T}\log_e\mathbb{E}_{\mathbf{z}}\{e^{-\theta TB
R}\}\,\quad \text{bits/s}. \label{eq:effectivedefirate}
\end{align}
Above, if  $R$ is equal to $R_0$ in (\ref{eq:commonrate}), then $C(\theta)$ is the maximum effective capacity (or equivalently the maximum constant arrival rate) of the common messages, which can be achieved with the power allocation policy $\mu = (\mu_0,\mu_1)$ under queueing constraints specified by the QoS exponent $\theta$. Similarly, if $R = R_1$ in (\ref{eq:inst-secrecyrate}), then $C(\theta)$ is the maximum effective capacity of the confidential messages achieved with the power control policy $\mu$. Note that in both cases, $R$ depends on the fading states $\mathbf{z}$, and the expectation in (\ref{eq:effectivedefirate}) is with respect to $\mathbf{z}$.

Finally, we denote the effective capacity normalized by bandwidth $B$
as
\begin{equation}\label{eq:normeffectivedefi}
\C(\theta)=\frac{C(\theta)}{B} \quad \text{bits/s/Hz}.
\end{equation}

\section{Effective Secure Throughput Region with Common Messages and Optimal Power Control} \label{sec:commonmessage}

In this section, we investigate the secure throughput region of and the optimal power control policies for the fading broadcast channel with
confidential messages (BCC) in the presence of statistical QoS
constraints. Hence, in the considered scenario, the transmitter sends common messages to two receivers, sends confidential messages to only one receiver, and operates under buffer constraints. Liang \emph{et al.} in \cite{liangsecure} showed
that the fading channel can be viewed as a set of parallel
subchannels with each subchannel corresponding to one fading state. Subsequently,
the ergodic secrecy capacity region is determined and the optimal
power allocation policies achieving the boundary of the capacity
region are identified in \cite{liangsecure}. Outage performance is also studied for cases in which long transmission delays cannot be tolerated and coding and decoding needs to be performed in one block.

\subsection{Effective Secure Throughput Region}

In this paper, we analyze the performance under statistical buffer
constraints by considering the effective capacity formulation.
Using the effective capacity expression in
(\ref{eq:effectivedefirate}), we first have the following definition for the effective throughput region.
\begin{Def}
The effective secure throughput region of the fading BCC is
\begin{align} \label{eq:securethroughputregion}
\mathcal{C}_{es}&=\bigcup_{\substack{\mathbf{R} = (R_0,R_1) \\ \text{s.t. } \,
\E\{\bf{R}\} \in \mathcal{C}_s}}\Bigg\{(\C_0,\C_1):
\C_j\le -\frac{1}{\theta TB}\log_e\E\{e^{-\theta T BR_j}\}\Bigg\}
\end{align}
where $\mathbf{R}=(R_0,R_1)$ is the vector composed of the
instantaneous rates for the common and confidential messages,
respectively.
\end{Def}

Note that the union in (\ref{eq:securethroughputregion}) is over the
distributions of the vector $\mathbf{R}$ such that the expected
value $\E\{\mathbf{R}\}$ lies in the ergodic secrecy capacity region
$\mathcal{C}_s$. Note also that the maximum values of the instantaneous rates $R_0$ and $R_1$ for a given power control policy $\mu$ are provided by (\ref{eq:commonrate}) and (\ref{eq:inst-secrecyrate}). Moreover, in (\ref{eq:securethroughputregion}), $C_0$ and $C_1$ denote the effective capacities of common and confidential messages, respectively.

Since the ergodic secrecy capacity region is convex as proved in Proposition
\ref{prop:convexity-ergodic}, we can easily prove the following.
\begin{Lem} \label{prop:convexity-effcap}
The effective secrecy throughput region $\mathcal{C}_{es}$ defined in (\ref{eq:securethroughputregion}) is convex.
\end{Lem}
\emph{Proof: }Let the two effective capacity pairs $\underline{\C}
= (\C_{0}, \C_{1})$ and $\underline{\C}' = (\C'_{0}, \C'_{1})$
belong to $\mathcal{C}_{es}$. Therefore, there exist some
$\mathbf{R} = (R_0, R_1)$ and $\mathbf{R'} = (R_0', R_1')$ for $\underline{\C}$
and $\underline{\C}'$, respectively. By a time sharing
strategy, for any $\alpha\in(0,1)$, we know that
$\E\{\alpha\mathbf{R}+(1-\alpha)\mathbf{R'}\}\in\mathcal{C}_{s}$. Then, we can write
\begin{align}
&\alpha\underline{\C}+(1-\alpha)\underline{\C}'\nonumber\\
&=-\frac{1}{\theta TB}\log_e\left(\E\left\{e^{-\theta
TB\mathbf{R}}\right\}\right)^\alpha\left(\E\left\{e^{-\theta
TB\mathbf{R'}}\right\}\right)^{1-\alpha} \label{eq:convexityproof1}\\
&=-\frac{1}{\theta TB}\log_e\left(\E\left\{\left(e^{-\theta
TB\alpha\mathbf{R}}\right)^{\frac{1}{\alpha}}\right\}\right)^\alpha
\left(\E\left\{\left(e^{-\theta
TB(1-\alpha)\mathbf{R'}}\right)^{\frac{1}{1-\alpha}}\right\}\right)^{1-\alpha}\label{eq:convexityproof2}\\
&\leq-\frac{1}{\theta TB}\log_e\E\left\{e^{-\theta TB\left(\alpha
\mathbf{R}+(1-\alpha)\mathbf{R'}\right)}\right\}. \label{eq:convexityproof3}
\end{align}
Above, in (\ref{eq:convexityproof1}) -- (\ref{eq:convexityproof3}), all algebraic operations are with respect to each component of the vectors. For instance, the expression in (\ref{eq:convexityproof1}) denotes a vector whose components are $\left\{\frac{1}{\theta TB}\log_e\left(\E\left\{e^{-\theta
TBR_j}\right\}\right)^\alpha\left(\E\left\{e^{-\theta
TBR_j'}\right\}\right)^{1-\alpha} \right\}$ for $j = 0,1$. The
inequality in (\ref{eq:convexityproof3}) follows from H\"{o}lder's inequality.
Hence, $\alpha\underline{\C}+(1-\alpha)\underline{\C}'$
lies in the \emph{throughput region}, showing the convexity. \hfill$\square$

Due to the convexity property, the points on the boundary surface of the effective throughput
region $(\C_0^*,\C_1^*)$ can be obtained by solving the following
optimization problem
\begin{align}\label{eq:optproblem}
\max_{\mu\in\mathcal{U}} \, \lambda_0\C_0+\lambda_1\C_1
\end{align}
where $\lambda=(\lambda_0,\lambda_1)$ is any vector in
$\mathfrak{R}^2_+$, and $\C_0$ and $\C_1$ are the maximum effective capacity values for a given power control policy $\mu$, i.e., they are the effective capacity values when the instantaneous service rates are the ones given in (\ref{eq:commonrate}) and (\ref{eq:inst-secrecyrate}).

\subsection{Optimal Power Control}

Having characterized the effective secure throughput region, we turn our attention to optimal power control. 
Note that due to the introduction of QoS constraints, the
maximization is over the effective capacities while the
service rates are limited by the instantaneous channel capacities.

Next, we derive the optimality conditions for the optimal power allocation $\mu^*$ that solves
(\ref{eq:optproblem}). As also provided in Section \ref{subsec:capacityregion},
the maximal instantaneous common message rate for a given power control policy $\mu$ is
\begin{align}
R_{0}=\left\{\begin{array}{ll}
\log_2\left(1+\frac{\gamma\mu_0(\mathbf{z})z_E}{1+\gamma\mu_1(\mathbf{z})z_E}\right),\,&\mathbf{z}\in\mathcal{Z}\\
\log_2\left(1+\mu_0(\mathbf{z})z_M\right),\,&\mathbf{z}\in\mathcal{Z}^c\end{array}\right..
\end{align}
Similarly, the maximal instantaneous confidential message (or equivalently secrecy)
rate is
\begin{align}
R_1=\left\{\begin{array}{ll}
\log_2\left(\frac{1+\mu_1(\mathbf{z})z_M}{1+\gamma\mu_1(\mathbf{z})z_E}\right),\,&\mathbf{z}\in\mathcal{Z}\\
0,\,&\mathbf{z}\in\mathcal{Z}^c\end{array}\right..
\end{align}
Now, using these instantaneous service rates $R_0$ and $R_1$ in the effective capacity expressions and recalling the average SNR constraint in (\ref{eq:avgpower}), we can express the Lagrangian of the convex optimization problem in (\ref{eq:optproblem}) as
\begin{align}
\mathcal{J}=&-\frac{\lambda_0}{\beta \log_e 2}\log_e
\Bigg(\int_{\mathbf{z}\in\mathcal{Z}}\left(1+\frac{\gamma\mu_0(\mathbf{z})z_E}{1+\gamma\mu_1(\mathbf{z})
z_E}\right)^{-\beta}p_{\mathbf{z}}(z_M,z_E)d\mathbf{z}
+\int_{\mathbf{z}\in\mathcal{Z}^c}\left(1+\mu_0(\mathbf{z})z_M\right)^{-\beta}p_{\mathbf{z}}(z_M,z_E)d\mathbf{z}\Bigg)\nonumber\\
&-\frac{\lambda_1}{\beta \log_e
2}\log_e\Bigg(\int_{\mathbf{z}\in\mathcal{Z}}\left(\frac{1+\mu_1(\mathbf{z})z_M}{1+\gamma\mu_1(\mathbf{z})
z_M}\right)^{-\beta}p_{\mathbf{z}}(z_M,z_E)d\mathbf{z}
+\int_{\mathbf{z}\in\mathcal{Z}^c}p_{\mathbf{z}}(z_M,z_E)d\mathbf{z}\Bigg)\nonumber\\
&-\kappa\left(\E_{\mathbf{z}\in\mathcal{Z}
}\{\mu_0(\mathbf{z})+\mu_1(\mathbf{z})\}+\E_{\mathbf{z}\in\mathcal{Z}^c}\{\mu_0(\mathbf{z})\}\right)
\end{align}
where $\beta=\frac{\theta TB}{\log_e2}$, $p_{\mathbf{z}}(z_M,z_E)$ is the joint distribution function of the fading states $\mathbf{z} = (z_M,z_E)$, and $\kappa \ge 0$ is the Lagrange multiplier. Next, we define $(\phi_0,\phi_1)$ as
\begin{align}
\phi_0&=\int_{\mathbf{z}\in\mathcal{Z}}\left(1+\frac{\gamma\mu_0(\mathbf{z})z_E}{1+\gamma\mu_1(\mathbf{z})
z_E}\right)^{-\beta}p_{\mathbf{z}}(z_M,z_E)d\mathbf{z}
+\int_{\mathbf{z}\in\mathcal{Z}^c}\left(1+\mu_0(\mathbf{z})z_M\right)^{-\beta}p_{\mathbf{z}}(z_M,z_E)d\mathbf{z}, \intertext{and}
\phi_1&=\int_{\mathbf{z}\in\mathcal{Z}}\left(\frac{1+\mu_1(\mathbf{z})z_M}{1+\gamma\mu_1(\mathbf{z})
z_E}\right)^{-\beta}p_{\mathbf{z}}(z_M,z_E)d\mathbf{z}
+\int_{\mathbf{z}\in\mathcal{Z}^c}p_{\mathbf{z}}(z_M,z_E)d\mathbf{z}.
\end{align}
Below, we derive the optimality conditions (that the optimal power control policies should satisfy) by differentiating the Lagrangian with respect to $\mu_0$ in regions $\mathcal{Z}^c$ and $\mathcal{Z}$ and with respect to $\mu_1$ in $\mathcal{Z}$, and making the derivatives equal to zero:
\begin{align}
& \hspace{-.5cm}1)\frac{\lambda_0}{\phi_0\log_e2}(1+\mu_0z_M)^{-\beta-1}z_M-\kappa=0 \quad \forall \mathbf{z} \in \mathcal{Z}^c \label{eq:2optcond1}\\
& \hspace{-.5cm}2)\frac{\lambda_0}{\phi_0\log_e2}\left(1+\frac{\gamma\mu_0z_E}{1+\gamma\mu_1
z_E}\right)^{-\beta-1}\frac{\gamma z_E}{1+\gamma\mu_1z_E}-\kappa=0 \quad \forall \mathbf{z} \in \mathcal{Z} \label{eq:2optcond2}\\
& \hspace{-.5cm}3)-\frac{\lambda_0}{\phi_0\log_e2}\left(1+\frac{\gamma\mu_0
z_E}{1+\gamma\mu_1 z_E}\right)^{-\beta-1}\frac{\mu_0(\gamma
z_E)^2}{(1+\gamma\mu_1
z_E)^2}
+\frac{\lambda_1}{\phi_1\log_e2}\left(\frac{1+\mu_1
z_M}{1+\gamma\mu_1z_E}\right)^{-\beta-1}\frac{z_M-\gamma
z_E}{(1+\gamma\mu_1z_E)^2}-\kappa=0 \quad \forall \mathbf{z} \in \mathcal{Z}\label{eq:2optcond3}
\end{align}
where (\ref{eq:2optcond1})-(\ref{eq:2optcond3}) are obtained by
evaluating the derivative of $\mathcal{J}$ with respect to $\mu_0$ when
$\mathbf{z} \in \mathcal{Z}^c$, $\mu_0$ when $\mathbf{z} \in
\mathcal{Z} $, and $\mu_1$ when $\mathbf{z} \in \mathcal{Z}$,
respectively. Whenever $\mu_0$ or $\mu_1$ turns out to have negative
values through these equations, they are set to 0.

We immediately note from (\ref{eq:2optcond1}) that when $\mathbf{z} \in \mathcal{Z}^c$ (i.e., when $z_M \le \gamma z_E$ and no confidential messages are transmitted), the optimal power control policy for the common messages can be expressed as
\begin{align}
\mu_0=\left[\frac{1}{\alpha_1^{\frac{1}{\beta+1}}z_M^{\frac{\beta}{\beta+1}}}-\frac{1}{z_M}\right]^+ \quad \forall \mathbf{z} \in \mathcal{Z}^c \label{eq:optmu_0inZ^c}
\end{align}
where $\alpha_1=\frac{\kappa\phi_0\log_e2}{\lambda_0}$. Also, when $\mathbf{z} \in \mathcal{Z}$, if no power is allocated for confidential messages and hence $\mu_1 = 0$, then we see from (\ref{eq:2optcond2}) that we can write the optimal $\mu_0$ as
\begin{align}
\mu_0=\left[\frac{1}{\alpha_1^{\frac{1}{\beta+1}}(\gamma z_E)^{\frac{\beta}{\beta+1}}}-\frac{1}{\gamma z_E}\right]^+ \quad \forall \mathbf{z} \in \mathcal{Z}.
\end{align}
A final remark about $\mu_0$ is the following. Noting that the term $\left(1+\frac{\gamma\mu_0z_E}{1+\gamma\mu_1
z_E}\right)^{-\beta-1}$ in (\ref{eq:2optcond2}) is less than 1 for $\mu_0 > 0$, we have $\frac{\lambda_0}{\phi_0\log_e2}\frac{\gamma z_E}{1+\gamma
\mu_1z_E} > \kappa$ when $\mu_0 > 0$. Equivalently, having $\frac{\lambda_0}{\phi_0\log_e2}\frac{\gamma z_E}{1+\gamma
\mu_1z_E} \le \kappa$ implies that $\mu_0 = 0$.

Regarding the optimal power control for the confidential messages, we have the following observations from the optimality conditions. We remark that when $\mu_0 = 0$, (\ref{eq:2optcond3}) becomes
\begin{equation}\label{eq:2subcond2}
\frac{\lambda_1}{\phi_1\log_e2}\left(\frac{1+\mu_1
z_M}{1+\gamma\mu_1z_E}\right)^{-\beta-1}\frac{z_M-\gamma
z_E}{(1+\gamma\mu_1z_E)^2}-\kappa=0
\end{equation}
from which the optimal $\mu_1$ can be computed.

When we have both $\mu_0 > 0$ and $\mu_1>0$, the optimal power allocations can be obtained by solving (\ref{eq:2optcond2}) and (\ref{eq:2optcond3}) simultaneously. In this case, a certain condition that depends only on $\mu_1$ can be obtained. By combining the equations in (\ref{eq:2optcond2}) and (\ref{eq:2optcond3}) and applying several straightforward algebraic manipulations, we get
\begin{align}\label{eq:2subcond4}
\frac{\lambda_1}{\kappa \phi_1 \log_e2}\left(\frac{1+\mu_1
z_M}{1+\gamma\mu_1z_E}\right)^{-\beta-1}\frac{z_M-\gamma
z_E}{(1+\gamma\mu_1z_E)^2}-\left(\frac{\lambda_0}{\kappa \phi_0
\log_e2}\frac{\gamma z_E}{1+\gamma \mu_1
z_E}\right)^{\frac{1}{\beta+1}}=0
\end{align}
which depends only on $\mu_1$. The positive solution $\mu_1 > 0$ of this equation provides the optimal power control policy for the confidential messages. Once optimal $\mu_1$ is determined, the optimal policy $\mu_0$ can be easily found from (\ref{eq:2optcond2}).

As seen in the above discussion, we have closed-form expressions for the optimal power control policy for the common messages in special cases (e.g., when $\mathbf{z} \in \mathcal{Z}^c$ or when $\mu_1 = 0$). On the other hand, the optimal power control policy for the confidential messages does not assume simple closed-form formulas even in special cases. Hence, optimal power control is in general determined through numerical computations. Making use of the optimality conditions in (\ref{eq:2optcond1}) -- (\ref{eq:2optcond3}) and the characterizations in (\ref{eq:optmu_0inZ^c}) through (\ref{eq:2subcond4}), we propose the following algorithm to obtain the optimal power adaptation policies. This algorithm is used in the numerical results presented in Section \ref{subsec:num-bcc}.

\begin{codebox}
\Procname{$\proc{Algorithm\ PC }(Power\ Control)$} \li Given
$\lambda_0,\lambda_1$, initialize $\phi_0,\phi_1$; \li
\label{eq:step-initializekappa2}Initialize $\kappa$; \li
\label{eq:step-determine2}Determine
 $\alpha_1=\frac{\kappa\phi_0\log_e2}{\lambda_0}$,
 $\alpha_2=\frac{\kappa\phi_1\log_e2}{\lambda_1}$;
  \li
  \If $z_M-\gamma z_E>0$
  \li
   \Then \If $z_M-\gamma z_E>\alpha_2$
  \li
  \Then Compute $\mu_1$ from (\ref{eq:2subcond2});
  \li
  \If $\mu_1>\frac{1}{\alpha_1}-\frac{1}{\gamma z_E}$ or $\gamma
z_E<\alpha_1$ \li \Then $\mu_0=0$;
 \li \Else \If
(\ref{eq:2subcond4}) returns positive solution \li \Then Compute
$\mu_0$ and $\mu_1$ from (\ref{eq:2optcond2}) and
(\ref{eq:2optcond3});\End
 \li \Else $\mu_1=0$,
 $\mu_0=\left[\frac{1}{\alpha_1^{\frac{1}{\beta+1}}(\gamma z_E)^{\frac{\beta}{\beta+1}}}-\frac{1}{\gamma
 z_E}\right]^+$;\End
 \li \Else $\mu_1=0$,
 $\mu_0=\left[\frac{1}{\alpha_1^{\frac{1}{\beta+1}}(\gamma z_E)^{\frac{\beta}{\beta+1}}}-\frac{1}{\gamma
 z_E}\right]^+$;\End
 \li \Else $\mu_1=0$,
 $\mu_0=\left[\frac{1}{\alpha_1^{\frac{1}{\beta+1}}z_M^{\frac{\beta}{\beta+1}}}-\frac{1}{z_M}\right]^+$;
 \End
 \li Check if the obtained $\mu_0$ and $\mu_1$ satisfy the average
power constraint with equality; 
\li \If not satisfied with equality
\li \Then update the value of $\kappa$ and return to Step
\ref{eq:step-determine2}; 
\li \Else move to Step
\ref{eq:step-check2};
\End \li \label{eq:step-check2}Evaluate
$\phi_0$ and $\phi_1$ with the obtained power control policies; \li
Check if the new values of $\phi_0$ and $\phi_1$ agree (up to a
certain margin) with those used in Step \ref{eq:step-determine2};
\li \If do not agree \li \Then update the values of $\phi_0$ and
$\phi_1$ and return to Step \ref{eq:step-initializekappa2}; \li
\Else declare the obtained power allocation policies $\mu_1$ and
$\mu_2$ as the optimal ones. \End
\end{codebox}

\subsection{Numerical Results} \label{subsec:num-bcc}

\begin{figure}
\begin{center}
\includegraphics[width=\figsize\textwidth]{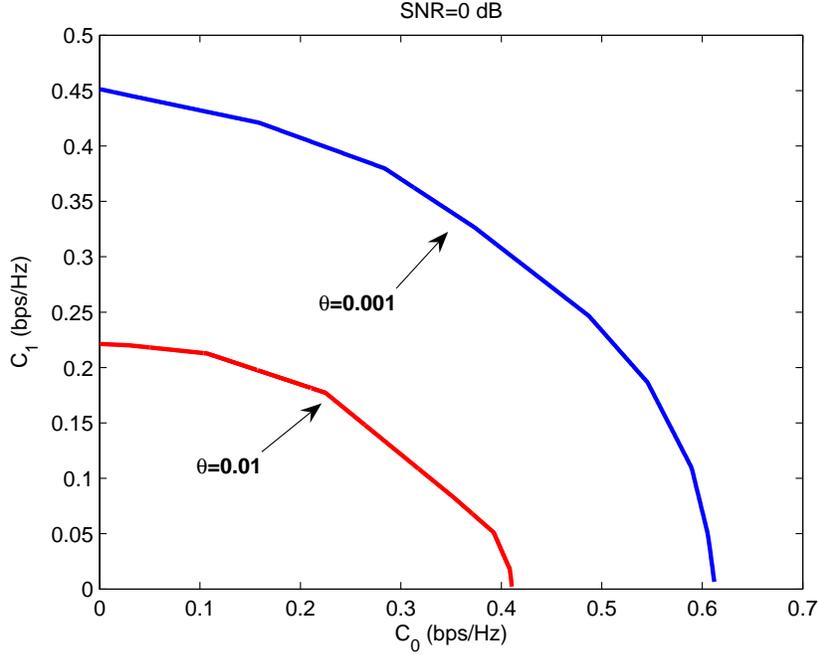}
\caption{The effective secrecy throughput region for $\theta=0.01$
 and $\theta=0.001$.}\label{fig:snr=1theta=01=001}
\end{center}
\end{figure}

In Fig. \ref{fig:snr=1theta=01=001}, we plot the
effective secrecy throughput region for
different $\theta$ values in a Rayleigh fading environment in which $z_M$ and $z_E$ are independent exponential random variables with $\E\{z_M\} = \E\{z_E\} = 1$.  We assume that $\gamma=1$, i.e., the
noise variances at both receivers are equal. We further
assume that $\tsnr=0$ dB. In the figure, we can observe that as $\theta$ increases and hence QoS
constraints becomes more stringent, the effective throughput region
shrinks. It is interesting to note that the percentage-wise decrease in the boundary point on the $y$-axis (i.e., the maximum effective secrecy capacity $\C_1$ when common message rate is
zero) is more than that in the boundary point on the $x$-axis (i.e., the maximum effective capacity $\C_0$ when confidential
message rate is zero). Hence, we see that the secrecy effective capacity is more severely affected by more strict QoS limitations.

\begin{figure}
\begin{center}
\includegraphics[width=\figsize\textwidth]{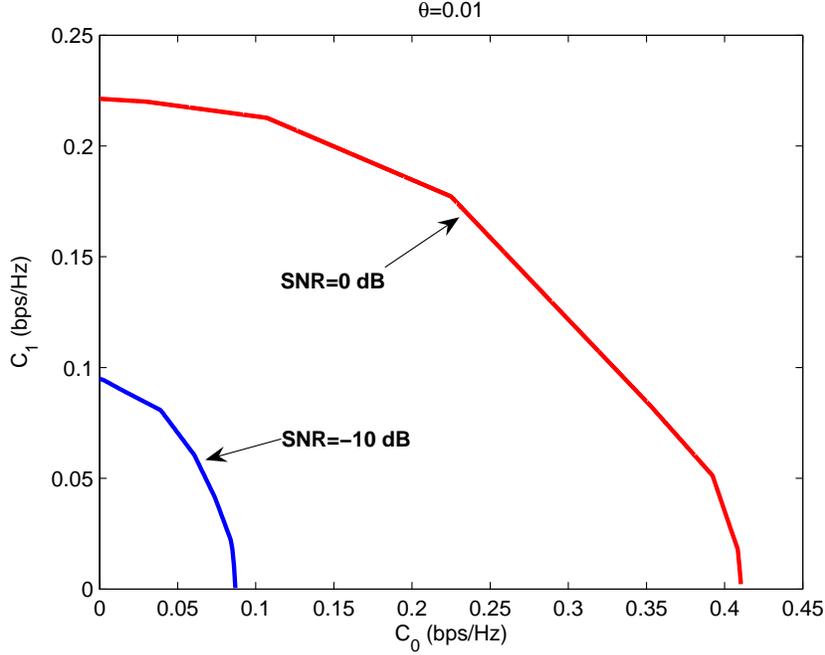}
\caption{The effective secrecy throughput region for $\tsnr=0$ dB
and $\tsnr=-10$ dB. $\theta=0.01$.}\label{fig:theta=01snr=1snr=01}
\end{center}
\end{figure}

We know that as $\tsnr$ decreases, the maximal
instantaneous service rate for common/confidential messages
decreases as well. In Fig. \ref{fig:theta=01snr=1snr=01}, we plot the
effective secrecy throughput region for different SNR values.
We assume that $\theta=0.01$. As we see from the figure, smaller
SNR introduces significant reduction in the effective throughput region.
Moreover, it is interesting to note that, contrary to the observation we had in Fig. \ref{fig:snr=1theta=01=001}, the decrease in the boundary point $\C_1$ when $\C_0=0$
is relatively smaller compared to the decrease in the boundary point $\C_0$ when
$\C_1=0$. Hence, a more severe impact is experienced by the common message rates.

\begin{figure}
\begin{center}
\includegraphics[width=\figsize\textwidth]{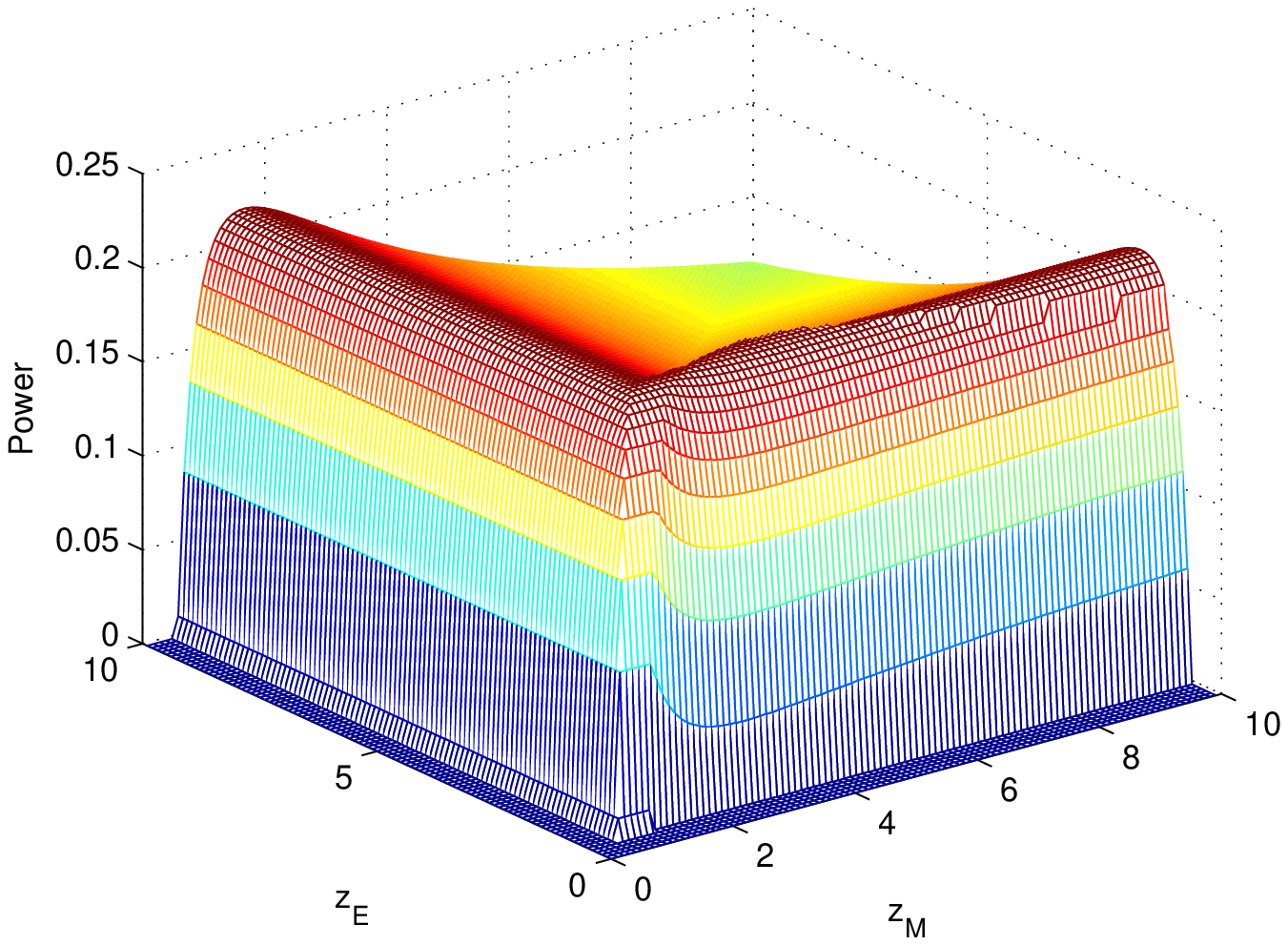}
\includegraphics[width=\figsize\textwidth]{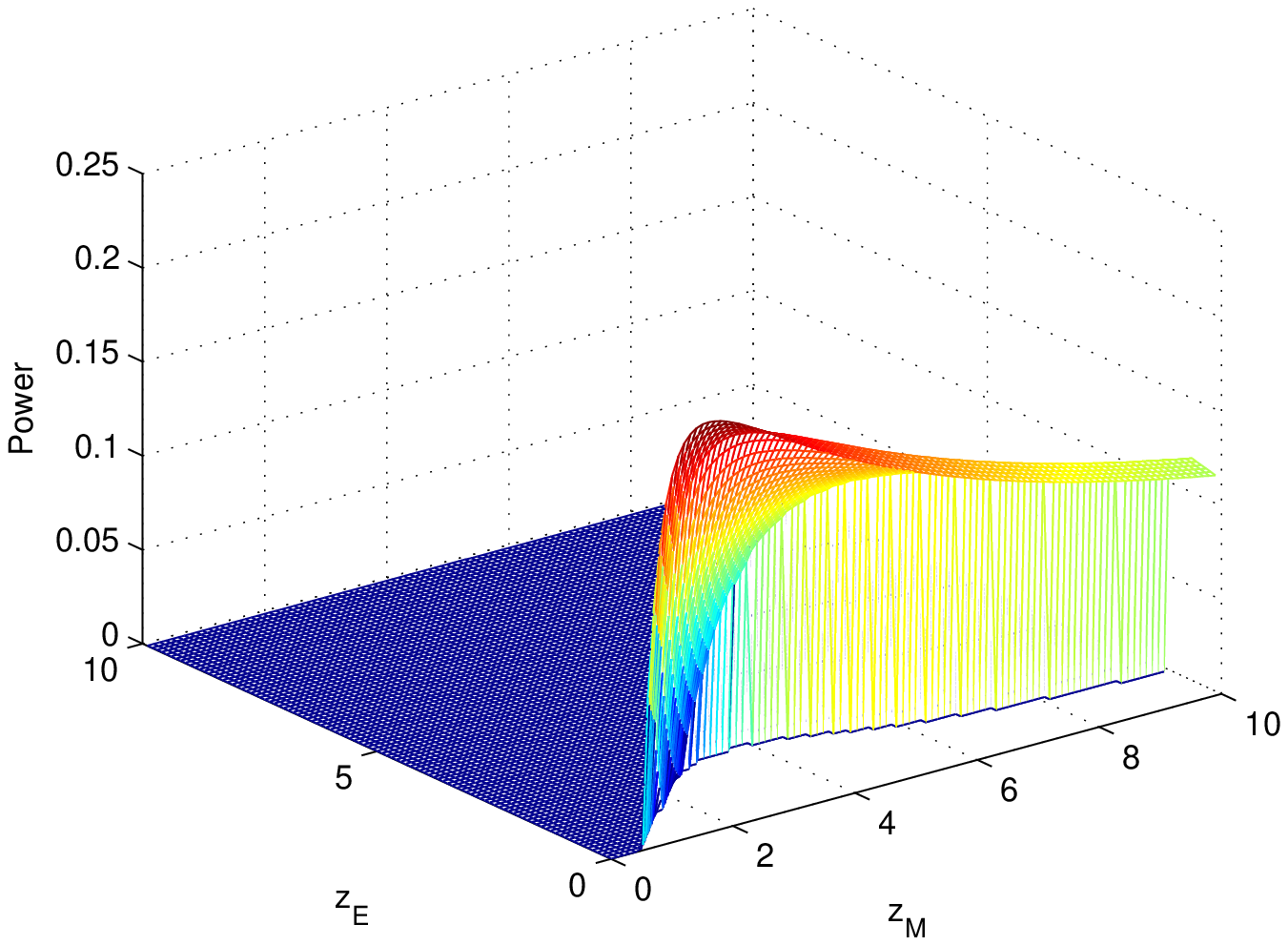}
\caption{The power allocated to common message and confidential
message transmissions as a function of $\mathbf{z} = (z_M, z_E)$. The top figure plots the power control policy
$\mu_0$, while the figure below plots the power control policy $\mu_1$.
$\lambda_0=0.5$. $\tsnr=-10$ dB. $\theta=0.01$.
}\label{fig:powerdistmu01}
\end{center}
\end{figure}

Finally, we plot the optimal power adaptation policies $\mu_0(\mathbf{z})$ and
$\mu_1(\mathbf{z})$ as a function of the channel states $\mathbf{z} = (z_M, z_E)$ in Fig.
\ref{fig:powerdistmu01}. In the figure, we have $\tsnr=-10$ dB and
$\theta=0.01$. Moreover, we assume $\lambda_0=0.5$, and hence these are the optimal power control policies that maximize the sum rate throughput $\C_0 + \C_1$. It is obvious from the figure
that the power for common message seems to be relatively uniformly distributed over
all the entire channel state space while the power for confidential messages
is concentrated in the smaller region $\mathcal{Z}$. Still, we note that the optimal $\mu_1$ provides relatively uniform distribution in $\mathcal{Z}$ rather than an opportunistic power allocation strategy in which more power is allocated to the transmission of confidential messages when $z_M$ is much larger than $z_E$ and less power otherwise. As will also be seen in the discussions of the following section, opportunistic power control is not necessarily optimal in the presence of buffer constraints as waiting until channel conditions gets favorable may lead to buffer overflows.

\section{Effective Secure Throughput and Optimal Power Control in the Absence of Common Messages} \label{sec:nocommonmessage}
In this section, we assume that the common message rate is zero, i.e., $R_0 = 0$, and
investigate the secrecy capacity and the associated optimal power control policy in the presence of QoS constraints. For this case, we identify equivalent optimization problems that are simpler to solve than the ones studied in Section \ref{sec:commonmessage}.
In particular, we analyze two types of power adaptation policies.
First, we consider the case in which the power control policies take
into account the CSI of both the main and eavesdropper channels.
Subsequently, we investigate power allocation strategies that are
functions of only the CSI of the main channel.

\subsection{Power Adaptation with Main and Eavesdropper Channel  State Information}\label{sec:fullcsi}

In this subsection, we assume that transmitter adapts the
transmitted power according to the instantaneous values of $z_M$ and
$z_E$. Recall that the instantaneous secrecy rate with power  adaptation policy $\mu(z_M,z_E)$ is given by
\begin{align}
R_1=\left\{\begin{array}{ll}
\log_2(1+\mu(z_M,z_E)z_M)
-\log_2(1+\gamma\mu(z_M,z_E)z_E),\,&\mathbf{z} \in \mathcal{Z}
\\0,\,&\mathbf{z} \in \mathcal{Z}^c \end{array}\right.
\end{align}
and
the maximum effective secure throughput can be expressed as
\begin{align}
\hspace{-1cm}\C_1&=\max_{\substack{\mu(z_M,z_E)
\\\E\{\mu(z_M,z_E)\}\le\tsnr}}-\frac{1}{\theta
TB}\log_e\Bigg(\int_0^\infty\!\!\!\!\int_0^{\gamma z_E}p_{z_M}(z_M)p_{z_E}(z_E)dz_Mdz_E\nonumber\\
&\hspace{-.4cm}+\int_0^\infty\int_{\gamma
z_E}^{\infty}\left(\frac{1+\mu(z_M,z_E)
z_M}{1+\gamma\mu(z_M,z_E)
z_E}\right)^{-\beta}p_{z_M}(z_M)p_{z_E}(z_E)dz_Mdz_E\Bigg)
\label{eq:fullcsimaxp}
\end{align}
where $p_{z_M}(z_M)$ and $p_{z_E}(z_E)$ are the probability density
functions of $z_M$ and $z_E$, respectively\footnote{In Section \ref{sec:nocommonmessage}, we assume that $z_M$ and $z_E$ are independent. While this is not necessarily required in Section \ref{sec:fullcsi}, optimal control policy results in Section \ref{subsec:maincsi} depend on this assumption. Hence, we have the same assumption throughout Section \ref{sec:nocommonmessage} for the sake of being consistent.}, and $\beta =
\frac{\theta T B }{\log_e 2}$. Note that the first term in the
$\log$ function is a constant and $\log$ is a monotonically
increasing function. Therefore, the maximization problem in
(\ref{eq:fullcsimaxp}) is equivalent to the following minimization
problem
\begin{align}
&\min_{\substack{\mu(z_M,z_E)\\\E\{\mu(z_M,z_E)\}\le\tsnr}}\int_0^\infty\int_{\gamma
z_E}^{\infty}\left(\frac{1+\mu(z_M,z_E)
z_M}{1+\gamma\mu(z_M,z_E)
z_E}\right)^{-\beta}
p_{z_M}(z_M)p_{z_E}(z_E)dz_Mdz_E. 
\label{eq:fullcsimaxprev}
\end{align}
It is easy to check that when $z_M>\gamma z_E$,
\begin{equation}
f(\mu)=\left(\frac{1+\mu z_M}{1+\gamma \mu z_E}\right)^{-\beta}
\end{equation}
is a convex function in $\mu$. Since nonnegative weighted sum of
convex functions is convex \cite{convex}, we can immediately see
that the objective function in (\ref{eq:fullcsimaxprev}) is also
convex in $\mu$. Then, we can form the following Lagrangian
function, denoted as $\mathcal{J}$:
\begin{align}
\mathcal{J}&=\int_0^\infty\int_{\gamma
z_E}^{\infty}\left(\frac{1+\mu(z_M,z_E) z_M}{1+\gamma
\mu(z_M,z_E)
z_E}\right)^{-\beta}p_{z_M}(z_M)p_{z_E}(z_E)dz_Mdz_E\nonumber\\
& +\lambda\left(\int_0^\infty\int_{\gamma
z_E}^{\infty}\mu(z_M,z_E)p_{z_M}(z_M)p_{z_E}(z_E)dz_Mdz_E-\tsnr\right).
\end{align}
Taking the derivative of the Lagrangian function with respect to
$\mu(z_M,z_E)$, we get the following optimality condition:
\begin{align}
&\frac{\partial \mathcal{J}}{\partial
\mu(z_M,z_E)}=\lambda-\beta\left(\frac{1+\mu(z_M,z_E)
z_M}{1+\gamma \mu(z_M,z_E)
z_E}\right)^{-\beta}
\frac{z_M-\gamma z_E}{(1+\mu(z_M,z_E)
z_M)(1+\gamma \mu(z_M,z_E) z_E)}=0\label{eq:fullcsioptcond}
\end{align}
where $\lambda$ is the Lagrange multiplier whose value is chosen to
satisfy the average power constraint with equality. For any channel
state pairs $(z_M,z_E)$, $\mu(z_M,z_E)$ can be obtained from
the above condition. Whenever the value of $\mu(z_M,z_E)$ is
negative, it follows from the convexity of the objective function
with respect to $\mu(z_M,z_E)$ that the optimal value of
$\mu(z_M,z_E)$ is 0.

There is no closed-form solution to (\ref{eq:fullcsioptcond}).
However, since the right-hand side (RHS) of
(\ref{eq:fullcsioptcond}) is a monotonically increasing function,
numerical techniques such as bisection search method can be
efficiently adopted to
derive the solution. 

The secure throughput can be determined by substituting the optimal
power control policy $\mu^*(z_M,z_E)$ in
(\ref{eq:fullcsimaxp}). Exploiting the optimality condition in
(\ref{eq:fullcsioptcond}), we can notice that when
$\mu(z_M,z_E)=0$, we have $z_M-\gamma
z_E=\frac{\lambda}{\beta}$. Meanwhile,
\begin{align}
&\left(\frac{1+\mu(z_M,z_E) z_M}{1+\gamma\mu(z_M,z_E)
z_E}\right)^{-\beta}
\frac{1}{(1+\mu(z_M,z_E) z_M)(1+\gamma
\mu(z_M,z_E) z_E)}<1.
\end{align}
Thus, we must have $z_M-\gamma z_E>\frac{\lambda}{\beta}$ for
$\mu(z_M,z_E)>0$, i.e., $\mu(z_M,z_E)=0$ if
$z_M-\gamma z_E\le\frac{\lambda}{\beta}$. Hence, we can write the
maximum effective secure throughput as
\begin{align}
\C_1&=-\frac{1}{\theta
TB}\log_e\Bigg(\int_0^\infty\int_0^{\gamma z_E+\frac{\lambda}{\beta}}p_{z_M}(z_M)p_{z_E}(z_E)dz_Mdz_E\nonumber\\
& +\int_0^\infty\int_{\gamma
z_E+\frac{\lambda}{\beta}}^{\infty}\left(\frac{1+\mu^*(z_M,z_E)
z_M}{1+\gamma \mu^*(z_M,z_E) z_E}\right)^{-\beta}
p_{z_M}(z_M)p_{z_E}(z_E)dz_Mdz_E\Bigg)
\end{align}
where $\mu^*(\theta,z_M,z_E)$ is the derived optimal power control
policy.

\subsection{Power Adaptation with  only Main Channel  State Information} \label{subsec:maincsi}

In this section, we assume that the transmitter adapts the power
level by only taking into account the CSI of the main channel (the
channel between the transmitter and the legitimate receiver). Under
this assumption, the instantaneous secrecy rate with power adaptation policy $\mu(z_M)$ is
\begin{align}
R_1=\left\{\begin{array}{ll}
\log_2(1+\mu(z_M)z_M)
-\log_2(1+\gamma\mu(z_M)z_E),\,&\mathbf{z} \in \mathcal{Z}
\\0,\,&\mathbf{z} \in \mathcal{Z}^c \end{array}\right.
\end{align}
and
the maximum effective secure throughput is
\begin{align}
\C_1&=\max_{\substack{\mu(z_M)\\\E\{\mu(z_M)\}\le\tsnr}}-\frac{1}{\theta
TB}\log_e\Bigg(\int_0^\infty\int_{z_M/\gamma}^{\infty}p_{z_M}(z_M)p_{z_E}(z_E)dz_Edz_M\nonumber\\
&\hspace{-.9cm}+\int_0^\infty\int_{0}^{z_M/\gamma}\left(\frac{1+\mu(z_M)
z_M}{1+\gamma \mu(z_M)
z_E}\right)^{-\beta}p_{z_M}(z_M)p_{z_E}(z_E)dz_Edz_M\Bigg)
. \label{eq:maincsimaxp}
\end{align}
Similar to the discussion in Section \ref{sec:fullcsi}, we get the
following equivalent minimization problem:
\begin{align}
\min_{\substack{\mu(z_M)\\\E\{\mu(z_M)\}\le\tsnr}}\int_0^\infty\int_{0}^{z_M/\gamma}&\left(\frac{1+\mu(z_M)
z_M}{1+\gamma \mu(z_M) z_E}\right)^{-\beta} 
p_{z_M}(z_M)p_{z_E}(z_E)dz_Edz_M
. \label{eq:maincsimaxprev}
\end{align}
The objective function in this case is again convex, and with a
similar Lagrangian optimization method, we can get the following
optimality condition:
\begin{align}
\frac{\partial \mathcal{J}}{\partial
\mu(z_M)}&=-\beta\int_{0}^{z_M/\gamma}\left(\frac{1+\mu(z_M)
z_M}{1+\gamma\mu(z_M)
z_E}\right)^{-\beta-1}
\frac{z_M-\gamma
z_E}{(1+\gamma\mu(z_M)
z_E)^2}p_{z_E}(z_E)dz_E+\lambda=0\label{eq:maincsioptcond}
\end{align}
where $\lambda$ is the Lagrange multiplier whose value is chosen to satisfy the average power
constraint with equality. If the obtained power level
$\mu(z_M)$ is negative, then the optimal value of
$\mu(z_M)$ becomes 0 according to the convexity of the
objective function in (\ref{eq:maincsimaxprev}). The RHS of
(\ref{eq:maincsioptcond}) is still a monotonic increasing function
of $\mu(z_M)$.

The secure throughput can be determined by substituting the optimal
power control policy $\mu^*(z_M)$ in (\ref{eq:maincsimaxp}).
Exploiting the optimality condition in (\ref{eq:maincsioptcond}), we
can notice that when $\mu(z_M,z_E)=0$, we have
\begin{align}
-\beta\int_{0}^{z_M/\gamma}(z_M-\gamma
z_E)&p_{z_E}(z_E)dz_E+\lambda=0
\\
&\Rightarrow \int_0^{z_M}P(z_E\le t/\gamma)dt=\frac{\lambda}{\beta}
\end{align}
Let us denote the solution to the above equation as $\alpha$.
Considering that
\begin{align}
\left(\frac{1+\mu(z_M) z_M}{1+\gamma \mu(z_M)
z_E}\right)^{-\beta-1}\frac{1}{(1+\gamma \mu(z_M) z_E)^2}<1,
\end{align}
we must have $z_M>\alpha$ for $\mu(z_M)>0$, i.e.,
$\mu(z_M)=0$ if $z_M\le\alpha$. Hence, we can write the maximum effective
secure throughput as
\begin{align}
\C_1&=-\frac{1}{\theta
TB}\log_e\Bigg(\int_0^{\alpha}\int_0^\infty
p_{z_M}(z_M)p_{z_E}(z_E)dz_Edz_M
+\int_{\alpha}^\infty
\int_{z_M/\gamma}^{\infty}p_{z_M}(z_M)p_{z_E}(z_E)dz_Edz_M\nonumber\\
&\hspace{2cm}+\int_{\alpha}^\infty
\int_0^{z_M/\gamma}\left(\frac{1+\mu^*(z_M) z_M}{1+\gamma
\mu^*(z_M) z_E}\right)^{-\beta}
p_{z_M}(z_M)p_{z_E}(z_E)dz_Edz_M\Bigg)
\end{align}
where $\mu^*(z_M)$ is the derived optimal power control
policy.


\subsection{Numerical Results}

\begin{figure}
\begin{center}
\includegraphics[width=\figsize\textwidth]{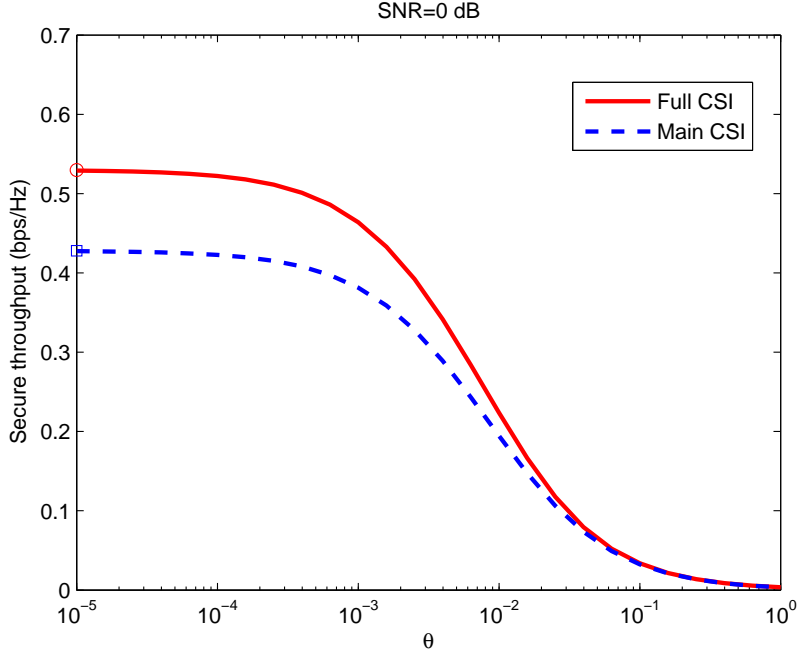}
\caption{The effective secure throughput vs. $\theta$ in the
Rayleigh fading channel with $\E\{z_E\}=\E\{z_M\}=1$. $\gamma=1$.
}\label{fig:secrecym=1}
\end{center}
\end{figure}

\begin{figure}
\begin{center}
\includegraphics[width=\figsize\textwidth]{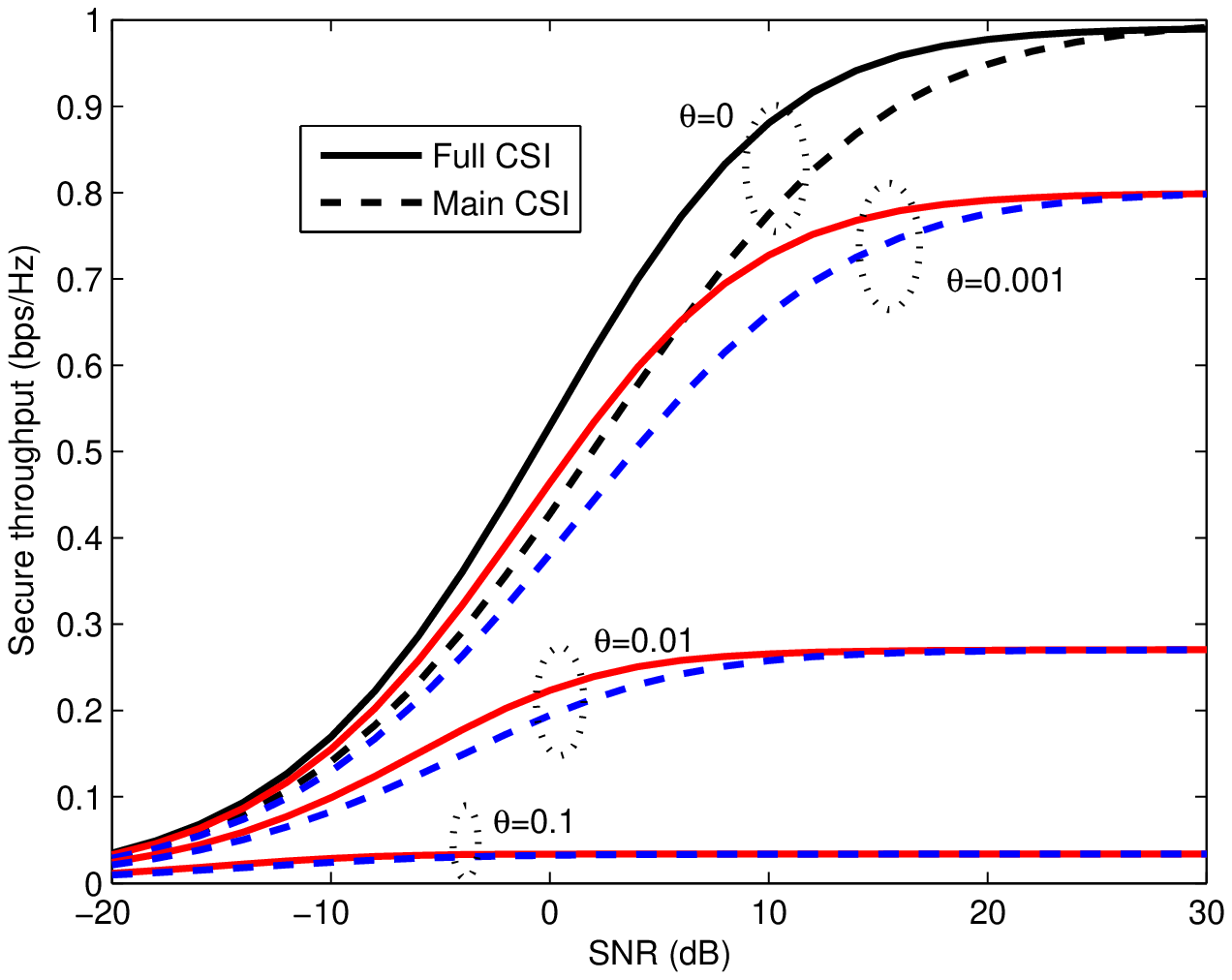}
\caption{The effective secure throughput vs. $\tsnr$ in the Rayleigh
fading channel with $\E\{z_E\}=\E\{z_M\}=1$.
$\gamma=1$.}\label{fig:fixedsnr}
\end{center}
\end{figure}

In Fig. \ref{fig:secrecym=1}, we plot the effective secure
throughput as a function of the QoS exponent $\theta$ in Rayleigh
fading channel with $\gamma=1$ when the power is adapted with
respect to the full CSI (i.e., the CSI of main and eavesdropper
channels) and also with respect to only the main CSI. It can be seen
from the figure that as the QoS constraints become more stringent
and hence as the value of $\theta$ increases, little improvement is
provided by considering the CSI of the eavesdropper channel in the
power adaptation. In Fig. \ref{fig:fixedsnr}, we plot the effective
secure throughput as $\tsnr$ varies for
$\theta=\{0,0.001,0.01,0.1\}$. Not surprisingly, we again observe
that taking into account the CSI of the eavesdropper channel in the
power adaptation policy does not provide much gains in terms of
increasing the effective secure throughput in the large $\tsnr$
regime. Also, as QoS constraints become more strict, we similarly
note that adapting the power with full CSI does not increase the
rate of secure transmission much even at medium $\tsnr$ levels.

\begin{figure}
\begin{center}
\includegraphics[width=\figsize\textwidth]{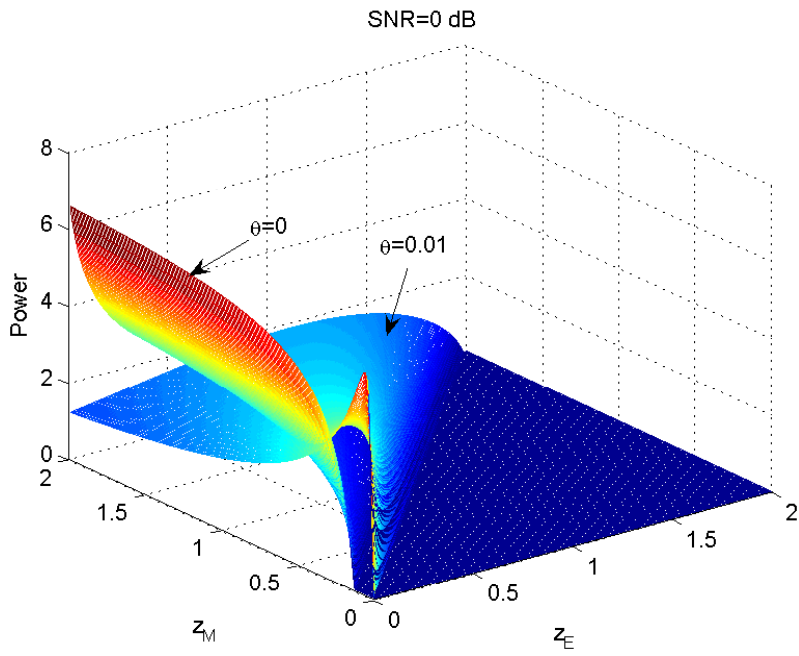}
\caption{The power allocation for the full CSI scenario with
$\tsnr=0$ dB in the Rayleigh fading channel with
$\E\{z_E\}=\E\{z_M\}=1$. $\gamma=1$. }\label{fig:powerdistfull-0}
\end{center}
\end{figure}

To characterize the power allocation strategy, we plot in Fig. \ref{fig:powerdistfull-0} the power
distribution as a function of $(z_M,z_E)$ for the full CSI case when
$\theta=0.01$ and $\theta=0$. In
the figure, we see that for both values of $\theta$, no power is
allocated for transmission when $z_M < z_E$ which is expected under
the assumption of equal noise powers, i.e., $N_1 = N_2$. We note
that when $\theta = 0$ and hence there are no buffer constraints,
opportunistic transmission policy is employed. More power is
allocated for cases in which the difference $z_M-z_E$ is large.
Therefore, the transmitter favors the times at which the main
channel is much better than the eavesdropper channel. At these
times, the transmitter sends the information at a high rate with
large power. When $z_M-z_E$ is small, transmission occurs at a small
rate with small power. However, this strategy is clearly not optimal
in the presence of buffer constraints because waiting to transmit at
a high rate until the main channel becomes much stronger than the
eavesdropper channel can lead to buildup in the buffer and incur
large delays. Hence, we do not observe this opportunistic
transmission strategy when $\theta = 0.01$. In this case, we note
that a more uniform power allocation is preferred. In order not to
violate the limitations on the buffer length, transmission at a
moderate power level is performed even when $z_M-z_E$ is small.

\section{Conclusion}
In this paper, we have investigated the fading broadcast channels
with confidential messages under statistical QoS constraints. We have
first defined the effective secrecy throughput region and proved the convexity of this region. Then, optimal power control policies that achieve the points on
the boundary of the throughput region are investigated. We have determined the conditions
satisfied by the optimal power control policies. In particular, we
have identified an algorithm for computing the optimal power allocated to
each fading state using the optimality conditions. When the broadcast
channel is reduced to the wire-tap channel with zero common message
rate, we have investigated two types of optimal power allocation
policies that maximize the effective secure throughput. In
particular, we have noted that the transmitter allocates power more
uniformly instead of concentrating its power for the cases in which
the main channel is much stronger than the eavesdropper channel. By
numerically comparing the obtained effective secure throughput, we
have shown that as QoS constraints become more stringent, the
benefits of incorporating the CSI of the eavesdropper channel in the
power control policy diminish.
\end{spacing}

\end{document}